\pdfoutput=1
\documentclass[aps,twocolumn, floatfix, prl]{revtex4}
\usepackage{graphicx}
\DeclareGraphicsExtensions{.pdf}
\usepackage{amsmath,amssymb,bbold,bm,color}
\usepackage{float}
\usepackage{epstopdf}

\newcommand{\bp}{{\bm p}}
\newcommand{\br}{{\bm r}}

\newcommand{\bsig}{{\bm \sigma}}
\newcommand{\btau}{{\bm \tau}}

\newcommand{\cC}{{\cal C}}

\newcommand{\cP}{{\cal P}}

\newcommand{\bee}{\begin{equation}}
\newcommand{\ee}{\end{equation}}

\begin{document}

\title{Proposed new platform to study interaction-enabled topological phases with fermionic particles}

\author{Ching-Kai Chiu}
\author{D.I. Pikulin}
\author{M. Franz}
\affiliation{Department of Physics and Astronomy, University of
British Columbia, Vancouver, BC, Canada V6T 1Z1}
\affiliation{Quantum Matter Institute, University of British Columbia, Vancouver BC, Canada V6T 1Z4}

\begin{abstract} 
We propose a new platform for interacting topological phases of fermions with time reversal symmetry $\bar{\Theta}$ (such that $\bar{\Theta}^2=1$) that can be realized in vortex lattices in the surface state of a topological insulator. The constituent particles are Majorana fermions bound to vortices and antivortices of such a lattice. We explain how the $\bar{\Theta}$ symmetry arises and discuss ways in which interactions can be experimentally tuned and detected.  We show how these features can be exploited to realize a class of  interaction-enabled crystalline topological phases that have no analog in weakly interacting systems.

\end{abstract}

\date{\today}

\maketitle

Theoretical understanding of topological phases of non-interacting fermions, now thought to be complete, gave us a treasure trove of new materials over the past decade including the topological insulators (both 2D and 3D) \cite{ReviewHasanKane,ReviewMoore,ReviewQiZhang,franz_book} and engineered topological superconductors with Majorana zero modes \cite{alicea_rev,beenakker_rev,stanescu_rev,elliot_rev}. The same theoretical framework also opened new areas of inquiry, leading to concepts such as the Weyl semimetals, axion insulators and topological exciton condensates that are actively pursued in experiments. When strong interactions are included the situation becomes more complicated and the classification of topological phases less well understood, especially for fermionic systems \cite{wen1}. It is expected that interacting systems could produce novel phases with unusual properties, such as excitations with fractionalized quantum numbers and unusual exchange statistics. Although a number of specific interaction-enabled topological phases with fermions have been theoretically proposed \cite{clarke1,lindner1,vaezi1,barkeshli1,wang1,metlitski1,bonderson1,chen1,mong1}, the only such phases that are known to exist in the real world are the fractional quantum Hall states \cite{tsui1,laughlin1}.  

With the goal of enlarging the space of experimentally accessible topological phases that depend for their existence on strong interactions we propose here a physical platform that realizes the paradigm of fermions with time reversal (TR) symmetry $\bar{\Theta}$ such that $\bar{\Theta}^2=1$. Interacting fermions with this property have been employed in seminal works by Fidkowski and Kitaev (FK) \cite{fidkowski1} who showed how the integer classification of their 1D topological phases in the non-interacting limit changes to Z$_8$ classification when interactions are included. 
More recently,  Lapa, Teo and Hughes (LTH) \cite{lapa1} introduced a model with $\bar{\Theta}$ and an additional inversion symmetry $\cP$. In this case, it turns out that there are no topologically non-trivial 1D phases in the absence of interactions. Remarkably, when interactions are included, a topologically non-trivial phase becomes possible (the interacting classification here is Z$_2$). The latter, then, is a genuine interaction-enabled topological phase that fundamentally cannot exist in the non-interacting limit. 

The TR symmetry  that we have in mind acts on the spinless fermion annihilation operator $c_j$ through an operator $\bar{\Theta}$ so that $\bar{\Theta} c_j\bar{\Theta}^{-1}=c_j$ and  $\bar{\Theta} i\bar{\Theta}^{-1}=-i$. If we decompose our Dirac fermion into a pair of Majorana fermions $c_j={1\over 2}(\alpha_j+i\beta_j)$ it follows that 
\begin{equation}\label{op1}
\bar{\Theta} \alpha_j\bar{\Theta}^{-1}=\alpha_j, \ \ \ \ 
\bar{\Theta} \beta_j\bar{\Theta}^{-1}=-\beta_j.
\end{equation}
In a $\bar{\Theta}$-invariant Hamiltonian expressed in terms of  $\alpha_j$, $\beta_j$ operators only certain terms are allowed. For instance bilinears $i\alpha_j\beta_k$ are allowed while $i\alpha_j\alpha_k$ and $i\beta_j\beta_k$ are prohibited. Similarly, four-fermion interaction terms with even numbers of $\alpha_j$'s are allowed, such as $\alpha_j\alpha_k\alpha_l\alpha_m$ or $\beta_j\beta_k\alpha_l\alpha_m$, but those with an odd number are not. Finding a physical system whose Hamiltonian implements the above symmetry constraints presents a challenge because electrons, the only relevant fermions in solids, have spin ${1\over 2}$ and the natural TR operation ${\Theta}$ for such spinfull fermions satisfies ${\Theta}^2=-1$. The important ideas \cite{fidkowski1,lapa1,turner1} that involve fermions with $\bar\Theta$ such that $\bar\Theta^2=1$ have therefore remained largely untested (see however proposal in Ref.\ \cite{fang1}).  In what follows we show how the ingredients necessary to realize the phases envisioned by FK and LTH can be realized in a concrete physical system accessible with the available experimental methods.

A system we explore in this work is similar to that studied in Ref.\ \cite{chiu1} -- a superconducting (SC) surface of a strong topological insulator (STI) -- with one extra ingredient. In addition to vortices, which are known to harbor unpaired Majorana zero modes (MZMs) \cite{fu1}, we include in our considerations antivortices which also contain MZMs but, as we show, of a different type. Specifically, we demonstrate that it is consistent to assign the two types of Majorana fermions  $\alpha_j$, $\beta_j$ that obey Eq.\ (\ref{op1}) to vortices and antivortices, respectively. Structures composed of vortices and antivortices, arranged such that their corresponding Majorana wavefunctions have non-zero overlaps, then implement $\bar\Theta$-invariant fermionic Hamiltonians
with $\bar\Theta^2=1$. We discuss various vortex/antivortex geometries that realize interacting lattice models, including the LTH model \cite{lapa1} mentioned above.

We now proceed to substantiate these ideas and claims. The physical system, an STI with a superconducting surface, is described by the Fu-Kane Hamiltonian \cite{fu1} ${\cal H}=\int d^2r \hat{\Psi}^\dagger_\br H_{\rm FK}(\br)\hat{\Psi}_\br,$ 
where
$\hat\Psi_\br=(c_{\uparrow\br},c_{\downarrow\br},c^\dagger_{\downarrow\br},-c^\dagger_{\uparrow\br})^T$
is the Nambu spinor and   
\begin{equation}\label{sym1}
H_{\rm FK}= \tau^z(\bp\cdot\bsig-\mu)+\tau^x\Delta_1+\tau^y\Delta_2.
\end{equation}
Here $\bsig$, $\btau$ are Pauli matrices in spin and Nambu spaces, respectively, and $\Delta=\Delta_1+i\Delta_2$ represents the SC order parameter.
A single isolated vortex, expressed as $\Delta(\br)=\Delta_0(r)e^{-in\varphi}$ with $\varphi$ the polar angle and $n=\pm 1$ corresponding to vortex or antivortex, respectively, is known to bind a MZM \cite{fu1}. 
The Hamiltonian (\ref{sym1}) respects the particle-hole symmetry $\cC$  generated by $\Xi=\sigma^y\tau^y K$ 
($\Xi^2=+1$, $K$ denotes complex conjugation) and, for a purely real gap function $\Delta$, also the physical TR symmetry $\Theta$ generated by  $\Theta=i\sigma^y K$ ($\Theta^2=-1$). In the presence of vortices $\Theta$  is broken but in the special case when $\mu=0$, the Hamiltonian  respects a fictitious TR symmetry $\bar{\Theta}$ with ${\bar\Theta}=\sigma^x\tau^x K$ ($\bar\Theta^2=+1$), even in the presence of vortices \cite{teo1}.
From now on we focus on this $\mu=0$ ``neutrality point'' with an extra symmetry.
Together, the two symmetries $\Xi$ and ${\bar\Theta}$ define a BDI class with chiral symmetry $\Pi=\Xi\bar{\Theta}=-\sigma^z\tau^z$. 

Eigenstates of the Hamiltonian (\ref{sym1}) are four-component Nambu spinors $\Phi(\br)=(u_{\uparrow},u_{\downarrow},v_{\uparrow},v_{\downarrow})^T$. Because the particle hole symmetry $\cC$ maps positive energy eigenstates to their negative energy partners, a non-degenerate zero mode $H_{\rm FK}\Phi_0=0$ is self-conjugate under $\cC$, that is, it obeys $\Xi\Phi_0=\Phi_0$. This constrains its components such that $v_{\uparrow}=u_{\downarrow}^*$ and $v_{\downarrow}=-u_{\uparrow}^*$. In addition, because $\{H_{\rm FK},\Pi\}=0$, the zero mode $\Phi_0$ must be an eigenstate of $\Pi$. There are two choices that satisfy these conditions, 
\begin{equation}\label{sym2}
\Phi_0^{(+)}=(0,u_{\downarrow},u_{\downarrow}^*,0)^T, \ \ \
\Phi_0^{(-)}=(u_{\uparrow},0,0,-u_{\uparrow}^*)^T,
\end{equation}
corresponding to the two eigenvalues $\nu=\pm 1$ of $\Pi$. It is easy to check that 
\begin{equation}\label{sym3}
\bar\Theta \Phi_0^{(\pm)}=\pm\Phi_0^{(\pm)},
\end{equation}
{\em i.e.}\ the two  MZMs transform as even and odd under $\bar\Theta$. To complete the argument we construct the corresponding zero mode operators $\gamma_{\pm}=\int d^2r\Phi_0^{(\pm)}(\br)^\dagger\hat\Psi_\br$. From Eqs.\ (\ref{sym2}) and (\ref{sym3}) it follows that 
\begin{equation}\label{sym4}
\gamma_\pm^\dagger=\gamma_\pm, \ \ \ \ \
\bar\Theta\gamma_\pm\bar\Theta^{-1} =\pm \gamma_\pm.
\end{equation}
The zero mode operators are Majorana and the two types transform in the opposite way under $\bar\Theta$. Now suppose that $\gamma_+$ with a wavefunction $\Phi_0^{(+)}$ resides in the core of a positive $n=1$ vortex (as can be verified by an explicit calculation \cite{chiu1}). The physical TR symmetry $\Theta$ maps a vortex onto an antivortex because the direction of superflow is reversed under $\Theta$.
At the same time the corresponding operator $\Theta$ maps $\Phi_0^{(+)}$ to $\Phi_0^{(-)}$. Antivortex thus necessarily carries the other type of Majorana represented by $\gamma_-$. Comparing Eqs.\ (\ref{sym4}) and (\ref{op1}) we conclude that vortices and antivortices in the Fu-Kane model at neutrality carry MZMs that transform as even and odd, respectively, under $\bar\Theta$ and can thus be assigned as $\alpha_j$, $\beta_j$ Majorana operators in models with $\bar\Theta^2=1$. We note that these results can be explicitly checked by a direct calculation of the zero-mode wavefunctions \cite{elliot_rev} in Hamiltonian (\ref{sym1}).

\begin{figure}[t]
\includegraphics[width = 8.5cm]{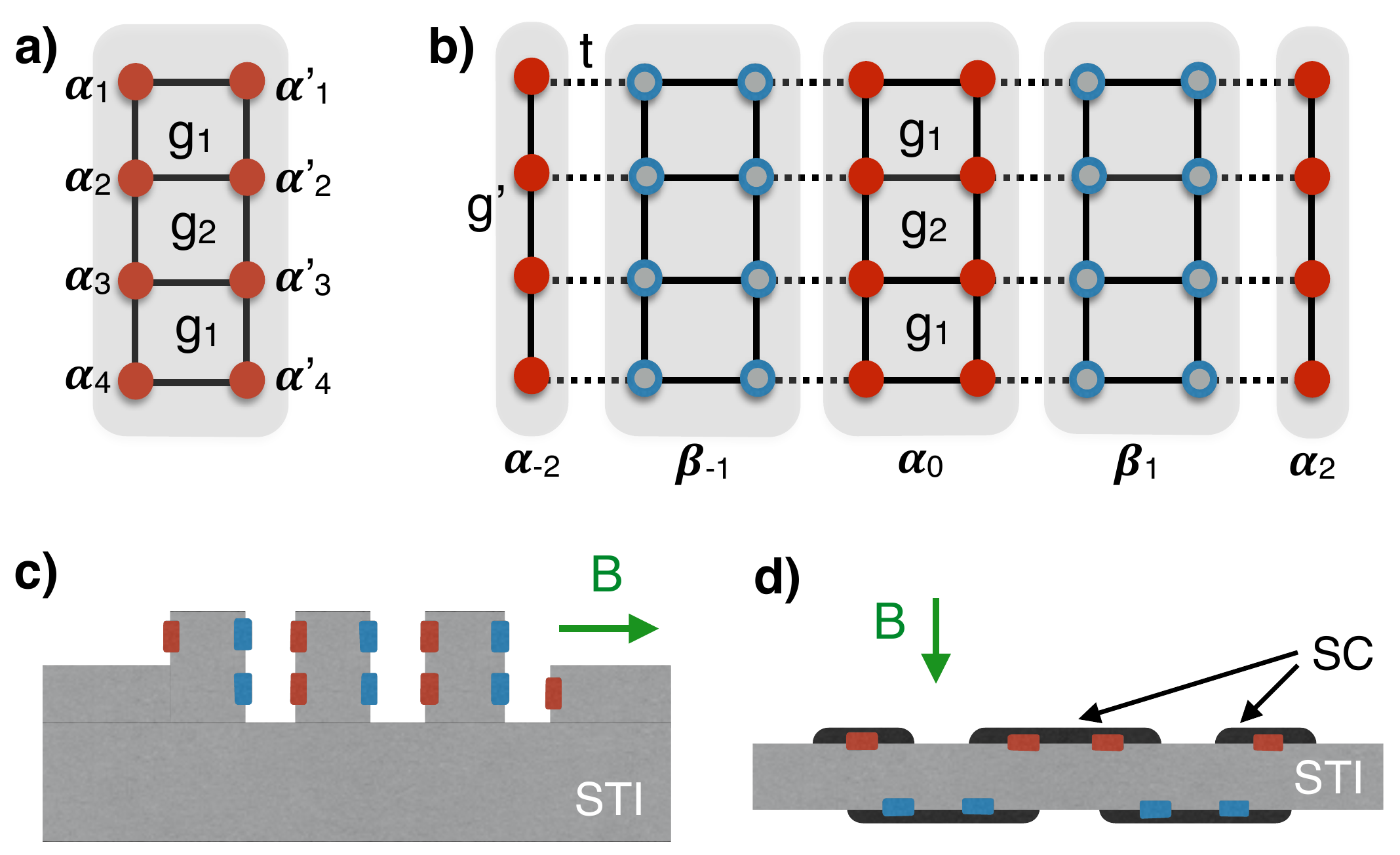}
\caption{Vortex lattice geometries for interacting Majorana models. Solid (open) circles represent vortices (antivortices), while solid and dashed lines indicate hopping and interaction terms, respectively, consistent with  $\bar\Theta$.  a) Cluster of 8 vortices. b) The 4-leg LTH ladder invariant under both $\bar\Theta$ and $\cP$.  Possible experimental realizations are sketched in panels c) and d). In panel (c) the STI surface is assumed to be superconducting and the thickness of the STI sufficiently large so that the Majorana wavefunction overlaps occur predominantly in the surface. Under these conditions the setup will realize the LTH ladder depicted in Fig.\ \ref{fig1}b. In panel (d) the interactions take place in the surfaces but tunneling is assumed to occur through the bulk of the flake.
}\label{fig1}
\end{figure}
Systems in this symmetry class composed of only vortices and no antivortices admit interaction terms of the form $g\alpha_j\alpha_k\alpha_l\alpha_m$ but no bilinear terms and are thus inherently strongly interacting. Some models of this type have been explored in Ref.\ \cite{chiu1}. When the system is slightly detuned from neutrality then $\bar\Theta$ is broken and bilinears $it'\alpha_j\alpha_k$ become allowed with $t'\sim\mu$. In the following we shall adopt an assumption that $\mu$ has been tuned sufficiently close to zero so that $t'\ll g$ and we may thus neglect all bilinears prohibited by $\bar\Theta$.  

\begin{figure}[t]
\includegraphics[width = 7.0cm]{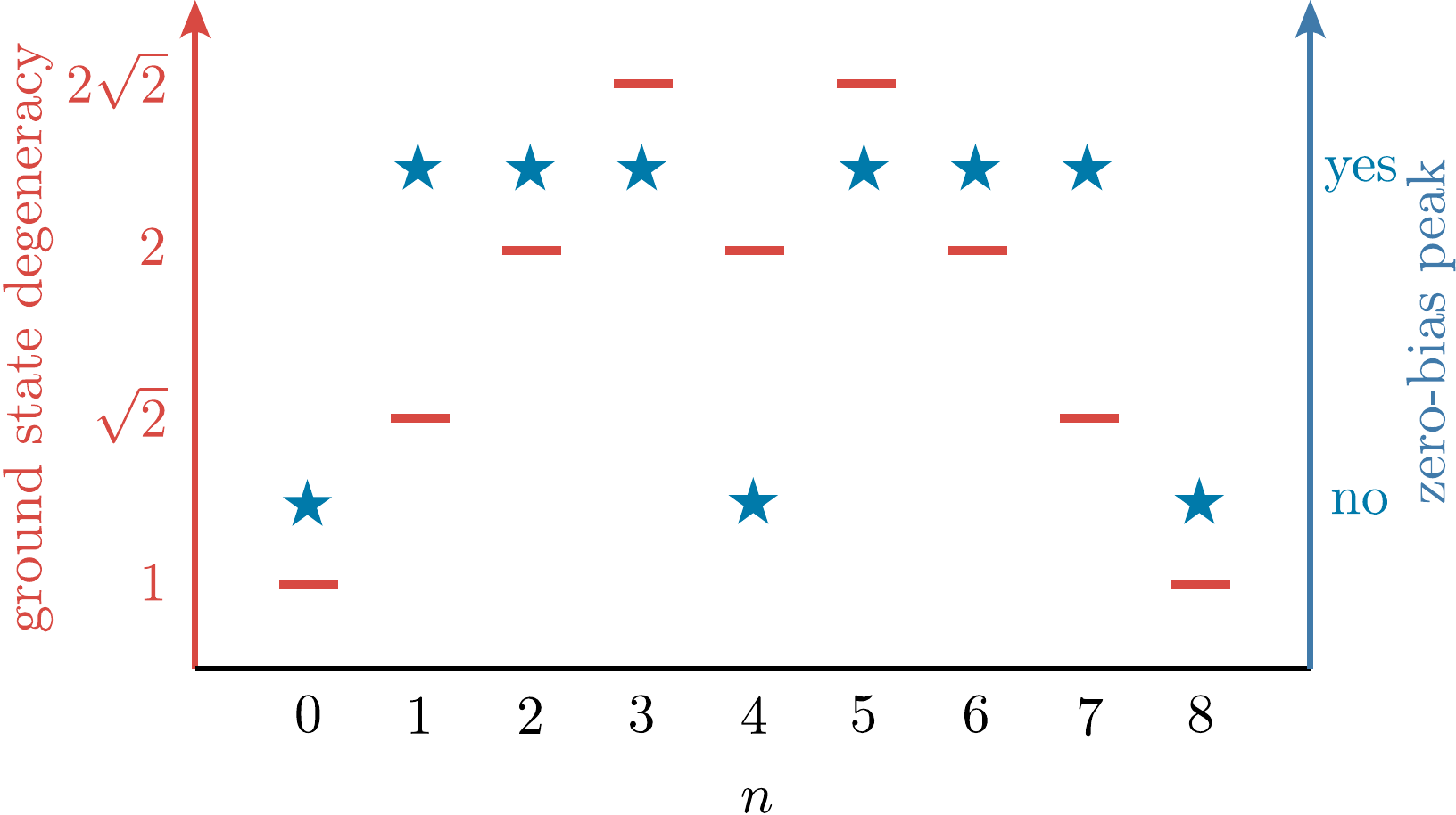}
\caption{Ground state degeneracy in a cluster with $n$ vortices and generic interaction terms. Stars denote presence or absence of zero bias peaks in vortices observable by STM. 
}\label{fig2}
\end{figure}

As our first example of interesting structures that can be constructed with these ingredients we consider a cluster made with a small number $n$  of vortices. As we shall argue, scanning tunneling microscopy (STM) can be used to probe the effects of interactions in such a small cluster by examining the vortex core spectra for the presence or absence of MZMs. 
 We envision changing the total vorticity $n$ of our cluster from 0 to 8 by adding vortices one by one to observe the theoretically predicted Z$_8$ periodicity \cite{fidkowski1,turner1} generated in the presence of interactions. In the absence of interactions the ground state has degeneracy $2^{n/2}$ and STM will observe a single MZM in each vortex. When interactions are present, STM should still see zero modes for $n=1,2,3$ and for all odd values of $n$, but the zero modes could generically split for even $n\geq 4$, because an interaction term can be first constructed with 4 Majorana operators. The pattern of ground state degeneracies we obtain for a system of $n$ vortices with generic interactions allowed by $\bar\Theta$ is displayed in Fig.\ \ref{fig2} and is consistent with results of Ref.\ \cite{turner1}. We find that MZMs, detectable by STM, will be present for all $n$ except when $n=4k$ with $k$ integer. This is to be contrasted with a non-interacting case with generic hopping terms allowed (e.g.\ when $\mu\neq 0$); here the ground state degeneracy is $\sqrt{2}(n\ {\rm mod}\ 2)$ and zero bias peaks will be seen for all odd $n$.

As an example of considerations that lead to Fig.\ \ref{fig2} we now discuss the special cases of an interacting system with $n=4k$. We start with
$n=4$, denote the Majorana operators as $\alpha_1$, $\alpha'_1$, $\alpha_2$, $\alpha'_2$, and define complex fermions $d_l={1\over 2}(\alpha_l+i\alpha_l')$. Noting that $i\alpha_l\alpha_l'=2n_l-1$ where $n_l=d^\dagger_l d_l$ is the number operator we may label the quantum states of the 4 Majorana fermions in the cluster by the eigenvalues $n_l=0,1$ as $|n_1 n_2\rangle$.
The most general interaction term $h_4=g \alpha_1\alpha_2\alpha'_1\alpha'_2$  splits the 4-fold degeneracy into an even parity ground state doublet  $|00\rangle$, $|11\rangle$ (for $g>0$) and odd parity excited states $|01\rangle$, $|10\rangle$. Because both ground states have the same parity single-electron tunneling will necessarily cause transitions to the excited states and the STM peaks will appear at energies $\pm 2g$, not zero.  

 For $n=8$ we consider a specific pattern displayed in Fig.\ \ref{fig1}a. This pattern is easy to analyze and also forms the basic building block for the interaction-enabled LTH topological crystalline phase that we shall discuss below. As argued previously \cite{chiu1} strongest interactions occur for those groups of 4 Majorana fermions that are spaced most closely together. This is because the corresponding coupling constants depend on the overlap of the exponentially decaying Majorana wavefunctions. With this in mind we identify the dominant interaction terms associated with the square plaquettes in Fig.\ \ref{fig1}a described by 
\begin{equation}\label{int1}
h_\square=g_1(\alpha_1\alpha_1'\alpha_2\alpha_2'+
\alpha_3\alpha_3'\alpha_4\alpha_4')+
g_2\alpha_2\alpha_2'\alpha_3\alpha_3'.
\end{equation}
 For $g_1,g_2>0$ the ground state of $h_\square$ is doubly degenerate, spanned  by eigenvectors $|0000\rangle$ and $|1111\rangle$ in the same notation as above. The energy is $E_g=-2g_1-g_2$.  
The subdominant interaction term is associated with two straight legs,
\begin{equation}\label{int2}
h_|=g'(\alpha_1\alpha_2\alpha_3\alpha_4+
\alpha_1'\alpha_2'\alpha_3'\alpha_4').
\end{equation}
It is easy to see that the inclusion of $h_|$ splits the ground state doublet of $h_\square$ into a bonding/antibonding pair with $|\psi_\pm\rangle=(|0000\rangle\pm |1111\rangle)/\sqrt{2}$. For $g'>0$ the unique ground state of 
\begin{equation}\label{int22}
h_8=h_\square+h_|
\end{equation}
is $|\psi_-\rangle$, while $|\psi_+\rangle$ is the first excited state with energy $4g'$. 

Now consider the 4-leg chain depicted in Fig.\ \ref{fig1}b. This is a version of the LTH model \cite{lapa1} that can be constructed using our platform. It consists of alternating clusters of 8 vortices and antivortices, each described by interacting Hamiltonian 
$h_8({\bm \alpha}_{2j})$  and $h_8({\bm \beta}_{2j+1})$, connected by nearest-neighbor hopping terms with amplitude $t$. Here ${\bm \alpha}_{2j}$ denotes the octet of $\{\alpha_l,\alpha_l'\}$ operators in cluster $2j$ and the same for ${\bm \beta}_{2j+1}$. The Hamiltonian, then, reads
\begin{eqnarray}\label{int3}
H_{\rm LTH}&=&\sum_{j=-N}^N h_8({\bm \alpha}_{2j})+ \sum_{j=-N-1}^Nh_8({\bm \beta}_{2j+1}) +h_|^{\rm edge}\nonumber\\
&-&it\sum_{j=-N-1}^{N+1}(\alpha_{l,2j}'\beta_{l,2j+1}+\beta_{l,2j-1}'\alpha_{l,2j}),
\end{eqnarray}
where $h_|^{\rm edge}$ has a form indicated in Eq.\ (\ref{int2}) and describes the four dangling Majoranas at the two ends of the chain. The Hamiltonian (\ref{int3}) respects $\bar\Theta$ as well as the inversion symmetry $\cP$. The latter is generated by $\{\alpha_{l,j},\alpha_{l,j}'\}\to\{\alpha_{l,-j}',\alpha_{l,-j}\}$ and $\{\beta_{l,j},\beta_{l,j}'\}\to-\{\beta_{l,-j}',\beta_{l,-j}\}$. 

\begin{figure}[t]
\includegraphics[width = 8.5cm]{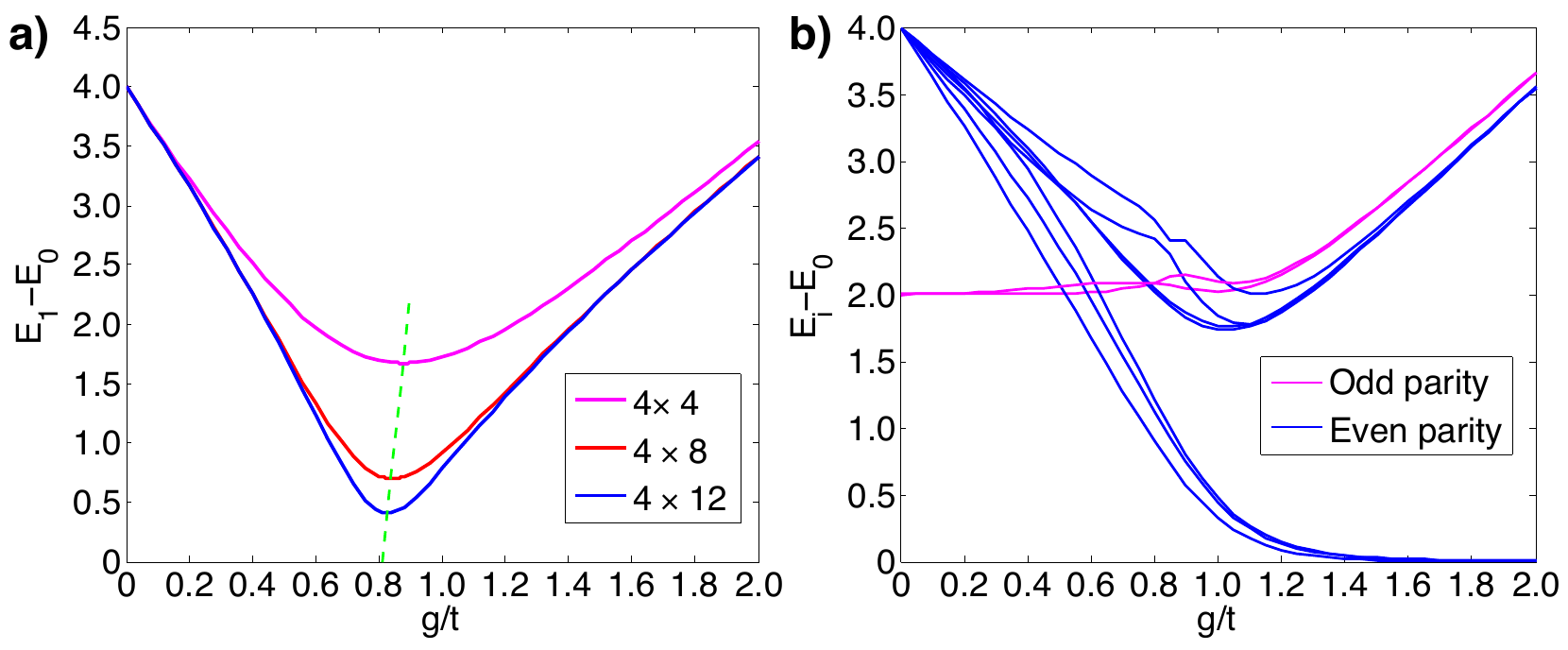}
\caption{Excitation energies of the LTH Hamiltonian (\ref{int3}) obtained by exact numerical diagonalization. a) The lowest excitation energy in a system of length $L_x=4,8,12$ with periodic boundary conditions.  b) Several lowest  excitation energies for $L_x=8$ and open boundary conditions. We take $g_1=g_2=g'\equiv g$ and the energies are in the units of $t$.
}\label{fig3}
\end{figure}
Without interactions, there exist no topologically non-trivial phases in a system with these symmetries \cite{lapa1}. Indeed we see that when $g_1=g_2=g'=0$  and $t\neq 0$ the system breaks up into a set of local dimers formed by $\alpha\beta$ products on the dashed bonds in Fig.\ \ref{fig1}b. The ground state is unique with a gap $2t$ to the lowest excitation and clearly topologically trivial.  

In the opposite limit, $t=0$ and $g_1\simeq g_2>g'>0$, the ground state is a direct product of $|\psi_-\rangle_j$ states on each cluster $j$. If we impose periodic boundary conditions the ground state is unique  with a gap $4g'$ to the lowest excited state. However, for an open chain that preserves the symmetries, as the one indicated in Fig.\ \ref{fig1}b, there is a {\em 4-fold degeneracy} associated with $h_|^{\rm edge}$. The four independent quantum states associated with the quartet of $\alpha_l$ Majoranas at each end are split by $h_|^{\rm edge}$ into a doubly degenerate ground state (with even local parity) and a doubly degenerate excited state (with odd parity). As noted in Ref.\ \cite{lapa1},  $\bar\Theta$ connects the two degenerate ground states, but its action in this subspace is {\em anomalous} with $\bar\Theta^2=-1$. The edge modes therefore comprise two effective spin-${1\over 2}$ degrees of freedom and constitute fractionalized excitations analogous to those appearing in spin-1 Haldane chains \cite{haldane1,aklt1}. 
In a long chain the edge degeneracy is therefore protected by the Kramers theorem and cannot be removed by any local perturbation preserving $\bar\Theta$.  It signals a topological phase, one that fundamentally cannot exist in a non-interacting system. If we turn on a small hopping $t$ the topological phase should persist up to a critical strength $t_c$ of order $g'$, at which point a phase transition occurs to the trivial phase.

We have confirmed the above picture by performing exact numerical diagonalizations of $H_{\rm LTH}$; the results are displayed in Fig.\ \ref{fig3}. Our simulations with periodic boundary conditions (panel a) indicate a phase transition marked by the excitation gap closing at $g_c\simeq
0.81 t$, obtained by extrapolating the energy minimum to $L_x\to\infty$. Panel b confirms that with open boundary conditions the ground state is unique for $g<g_c$ but becomes 4-fold degenerate for $g>g_c$ with all ground states in the same parity sector, in accord with our expectation for the interaction-enabled topological phase.

As explained in the Supplementary Material \cite{suppl1} a system described by the LTH Hamiltonian (\ref{int3}) can be regarded as a ``4$e$ superconductor''. (This is because it breaks the fermion number conservation symmetry while all the anomalous expectation values that are bilinear, such as $\langle d_1d_2\rangle$, vanish.) 
The authors of Ref.\ \cite{lapa1} proposed to look for physical realizations of such a 4$e$ superconductor in certain pair density wave systems, discussed theoretically in the context of underdoped high-$T_c$ cuprate superconductors \cite{berg1}.
Whether or not such a phase exists in real materials remains to be seen.
Our proposed realization of the LTH model, on the other hand, relies on ingredients that are known to exist. Specifically, SC order in the surface state of an STI has been experimentally observed by a number of groups \cite{koren1,sacepe1,wang11,qu1,williams1,fedorov1,sou1,cho1,elbaum1,xu0,xu1}. In some cases the chemical potential has been tuned to the close vicinity of the Dirac point  \cite{cho1,elbaum1} as required for models with symmetry $\bar\Theta$. Vortex cores \cite{xu1}, and the expected MZMs have also been imaged \cite{xu2}. Assembling vortices into regular structures, controlling their geometry and reliably probing the zero modes remains a challenge but it is one that does not seem insurmountable given the recent progress. 

In the near term it should be possible to probe the effects of interactions in small clusters of vortices by STM, as we discussed. The relevant energy scale can be quite large, $\sim 10$meV, under favorable conditions \cite{chiu1} and should permit observation of the interaction induced Z$_8$ periodicity.  To engineer the LTH model the key challenge will be to assemble stable arrays containing both vortices and antivortices in close proximity to one another. We note that such structures have been observed to occur spontaneously in mesoscopic SC samples with certain geometries \cite{chibotaru1}. Alternately, one can leverage the fact that in a thin STI/SC film or flake a uniform perpendicular field $B$ will produce vortices on the top surface and antivortices on the bottom surface. 
Panels (c) and (d) in Fig.\ \ref{fig1} outline two possible experimental realizations of the LTH vortex/antivortex lattice exploiting this principle
\cite{LTH2}.  

The interaction-enabled topological phases discussed in this work are gapped and therefore robust with respect to moderate amounts of symmetry-preserving disorder. Specifically, disorder in vortex positions does not break $\bar\Theta$ and is therefore innocuous. It also follows that fine details of the geometry sketched in Fig.\ \ref{fig1}(a,b) are unimportant as long as $\bar\Theta$ and $\cP$ are preserved to a good approximation. Any experimental realization that reasonably approximates the proposed model geometry should  show the topological phase.  Local fluctuations in the chemical potential break $\bar\Theta$ but we expect the gapped phases to remain robust as long as the symmetry is preserved on average and the fluctuations do not exceed the gap amplitude \cite{suppl1,cobanera1}.

\begin{acknowledgments}
The authors are indebted to  NSERC, CIfAR and Max Planck - UBC Centre for Quantum Materials for support. M.F.\ acknowledges The Aspen Center for Physics and IQMI at Caltech for hospitality during the initial stages of this project.
\end{acknowledgments}



\newpage
\section{Supplemetary Material}
\subsection{Disorder effects}

As mentioned in the main text we expect symmetry-preserving disorder to have only a mild effect on the gapped topological phase of the LTH model. Here we present numerical evidence that supports this expectation. We note that the roles played by the two symmetries $\bar\Theta$ and $\cP$ are somewhat different. The former underlies the existence of the topological phase and protects the exact degeneracy of its edge modes. The latter plays a subsidiary role by  prohibiting the existence of the topological phase in the absence of interactions. Consequently, we expect $\bar\Theta$-breaking disorder to have more pronounced effects on the system than the $\cP$-breaking disorder. This is indeed borne out by numerical simulations presented below.

To simulate the $\cP$-breaking disorder we replace the hopping  $t\to t+\delta t$ and the interaction $g\to g+\delta g$ in the LTH Hamiltonian Eq.\ (9). Here $\delta t$ is a random variable drawn independently on each bond from a uniform distribution $\delta t\in (-w_t,w_t)$ and similarly for  $\delta g\in (-w_g,w_g)$. Fig.\ 4 shows the lowest excitation energies of the disordered system with open boundary conditions.  We observe that the gap and the expected 4-fold ground state degeneracy in the topological phase remain robust for moderate levels of $\cP$-breaking disorder (top panels in Fig.\ 4). To simulate disorder that breaks $\bar\Theta$ we add  $\delta H$ to the LTH Hamiltonian Eq.\ (9) that contains terms $i\delta t'\alpha_{l,j}\alpha'_{l,j}$ and  $i\delta t'\beta_{l,j}\beta'_{l,j}$ describing tunneling events between Majorana sites within each cluster. Such terms are prohibited by $\bar\Theta$. The amplitude is again drawn from a random distribution  $\delta t'\in (-w_t',w_t')$. The results, displayed in bottom panels of Fig.\ 4, show that while the gap remains robust with respect to moderate levels of $\bar\Theta$-breaking disorder the exact degeneracy of the edge states is now lifted  in proportion to the disorder strength. This is akin to the gap formation observed in the surface state of a topological insulator induced by doping with magnetic impurities. Importantly, the edge modes remain nearly degenerate and clearly distinguishable from the bulk modes even for a fairly large disorder strength $w_t'=0.1t$. 

\begin{figure}[t]
\includegraphics[width = 8.5cm]{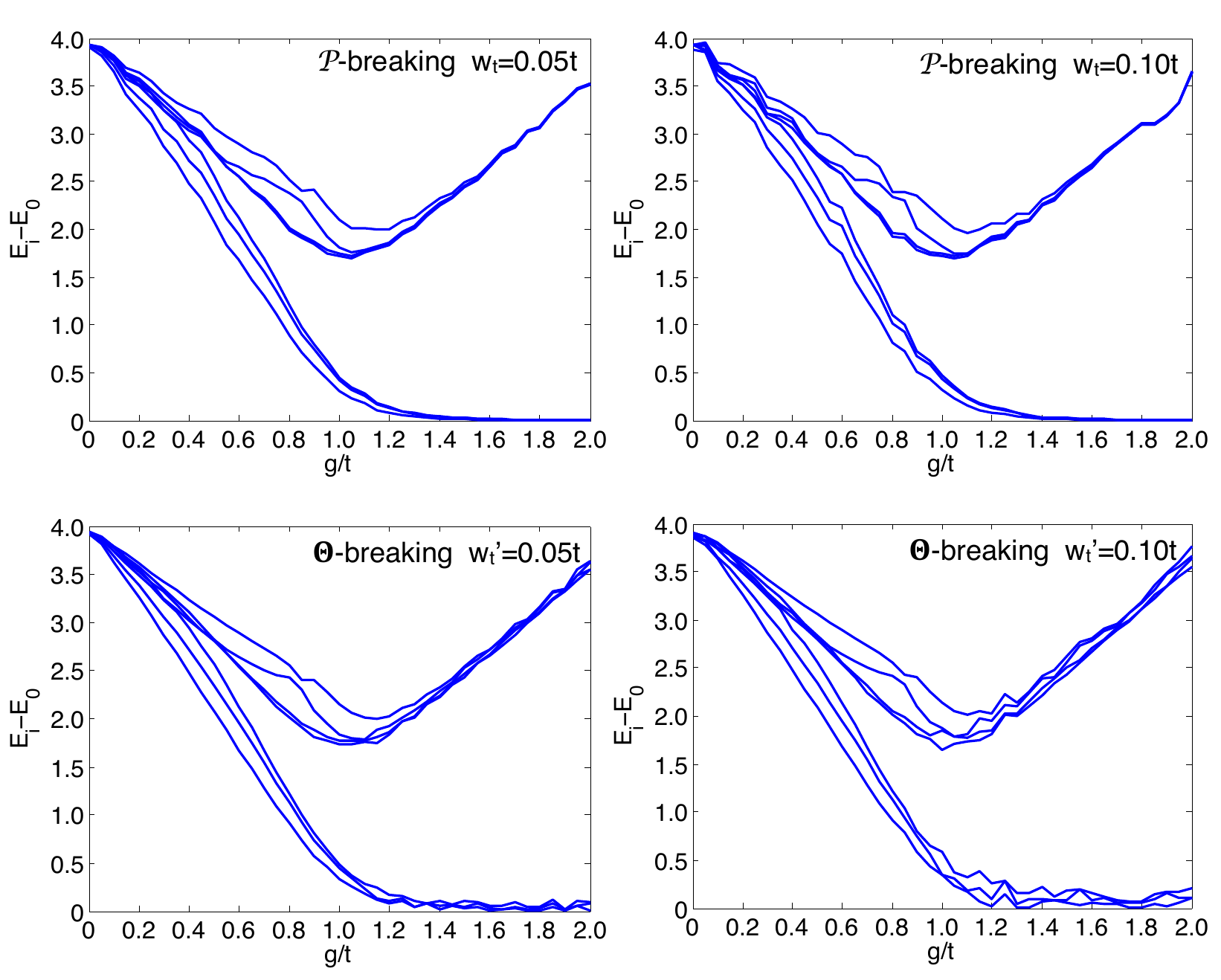}
\caption{Excitation energies of the LTH Hamiltonian (\ref{int3}) with disorder obtained by exact numerical diagonalization in a system of length $L_x=8$ with open boundary conditions. The same model parameters as in Fig.\ (3) are taken. The top row shows results for $\cP$-breaking disorder with strength $w_t=w_g=0.05t$ and  $w_t=w_g=0.10t$. The bottom row shows results for $\bar\Theta$-breaking disorder with strength $w_t'=0.05t$ and  $w_t'=0.10t$. Because these simulations are numerically costly we do not average over disorder realizations. However, each point in these graphs corresponds to an independent disorder realization which provides a clear sense for the statistical properties of the system. 
}\label{fig4}
\end{figure}
We conclude that for all practical purposes the interaction-enabled topological phase survives the inclusion of moderate levels of both $\bar\Theta$- and $\cP$-breaking disorder as expected on the basis of general arguments. The bulk gap and the gapless edge modes remain experimentally observable in the presence of disorder.  In fact these signatures of the phase should survive until the disorder strength becomes comparable to the bulk gap.

\subsection{Comments on the LTH model realization}
Fig.\ 1(b) represents an ideal configuration for the realization of the physics embodied by the LTH model. Some experimental geometries that can be used to approximate this configuration are sketched in panels (c,d) of that Figure. In these, red (blue) dots represent vortices (antivortices) and it is assumed that the lateral extent of both structures is such that it accomodates 4 vortices and antivortices in the direction perpendicular to the plane of the drawing. 

Aside from the dominant couplings $(g_1,g_2,g',t)$ considered in the model Hamiltonian Eq.\ (9) in the real physical system  all other couplings allowed by symmetry will also be present. For instance there will be additional hoppings  across the plaquette diagonal as well as additional four-fermion terms within each cluster and those involving two $\alpha$ and two $\beta$ operators between the clusters. Although not explicitly included in our numerical calculations, such symmetry-preserving terms will not alter the conclusions regarding the interaction-enabled topological phase in this model, as long as they remain small and short ranged. The latter property is guaranteed by the exponential decay of MZM wavefunctions while the former must be built into the geometry of the structure. The key point here is that {\em a generic} set of four-fermion interaction terms acting within a cluster of 8 MZMs produces a non-degenerate ground state, except for accidental degeneracies that occur on a set of measure zero in its parameter space. This is, ultimately, all we need to guarantee the existence of the gapped topological phase in the limit of strong interactions. Inclusion of the symmetry preserving hopping terms (again, of an arbitrary form) will then sustain the gapped phase for some range of hopping strength as illustrated in Figs.\ 3 and 4. 
To further document this robustness we have performed numerical computations using the LTH Hamiltonian (9) for different sets of model parameters. As an example we show in Fig.\ 5 the energy spectrum of the system for the case when $g'$ is smaller than both $g_1$ and $g_2$. As expected on general grounds this leads to the same qualitative results for the phase diagram of the system. 
We also note that, for the model to describe spinless fermions with time reversal symmetry, we require equal numbers of $\alpha$- and $\beta$-type MZMs as well as certain notion of locality, i.e.\ that MZM components of the complex fermion $c_j$ defined above Eq.\ (1) be not too far removed from one another. 
\begin{figure}[t]
\includegraphics[width = 4.5cm]{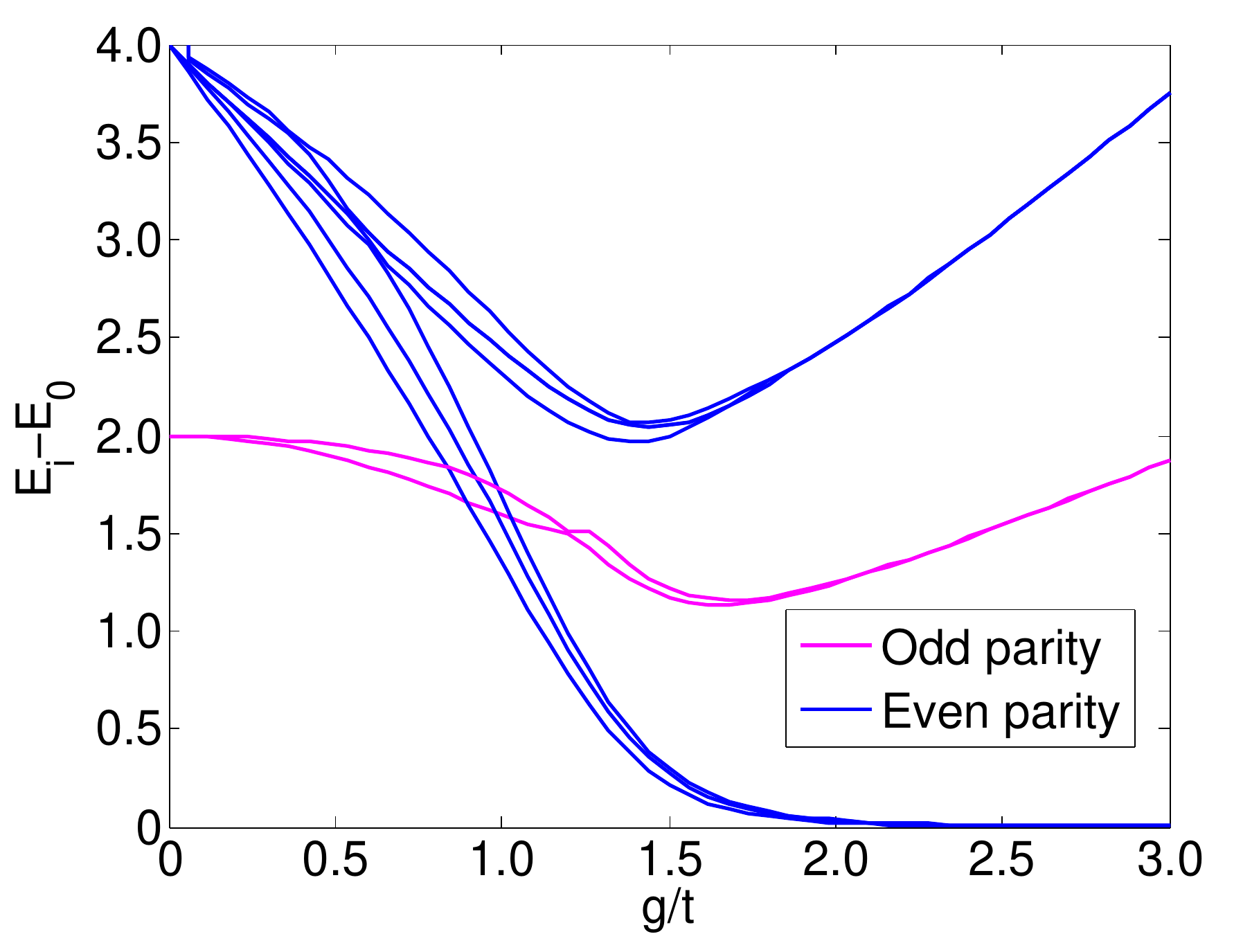}
\caption{Excitation energies of the LTH Hamiltonian (\ref{int3}) as a function of $g=g_1=g_2=3g'$.
}\label{fig5}
\end{figure}

These observations, combined with the robustness against disorder demonstrated above, indicate considerable leeway in the actual experimetal implementation. The experimental system need not have the precise geometry indicated in Fig.\ 1(b); so long as it consists of alternating clusters of 8 vortices and antivortices, connected with one another predominantly by tunneling terms, and so long as it preserves the $\cP$ and $\bar\Theta$ symmetries to a reasonable approximation, the interaction-enabled topological phase should naturally obtain in the regime when interactions dominate over the tunneling terms. 

While the geometry is rather flexible there are some basic requirements on the parameters of the physical system that must be met for it to realize the interacting phase. Most importantly, we require that $|\mu|$ is sufficiently close to zero so that the direct tunneling $t'\sim \mu$ between MZMs of the same type is negligible compared to the interaction scale $g$. We have previously estimated $g\simeq 10$meV under favorable conditions which means that $\mu$ must be tuned to zero to within few meV. Second, we require that MZMs are well separated in energy from all other vortex core states. The lowest lying such state has the energy $E_1\approx \Delta^2/\sqrt{\Delta^2+\mu^2}$. We see that when $\mu\ll\Delta$ this energy approaches the full SC gap so the second condition ultimately also boils down to the chemical potential being as small as possible while simultaneously maintaining significant gap amplitude $\Delta$.

\subsection{LTH model as an emergent $4e$ superconductor}
As mentioned in the main text the gapped topological phase of the LTH model can be regarded as a $4e$ superconductor (SC). Here we explain this notion in greater detail. First, the system should be considered a superconductor because its Hamiltonian (9) breaks the particle number conservation symmetry (although it conserves the fermion parity). In an ordinary SC such number non-conservation is associated with a non-vanishing anomalous expectation value $\Delta_{ll'}^{(2)}=\langle d^\dagger_l d^\dagger_{l'}\rangle$ often identified as the SC order parameter. Here, $d_l={1\over 2}(\alpha_l+i\alpha_l')$ is a complex fermion operator defined in the main text. However, deep in the topologial phase of the LTH model all such anomalous expectation values vanish. This can be seen most easily by considering the limit $g_1=g_2\gg g'$ and $t=0$ of the model. As discussed below Eq.\ (7) the ground state in this limit can be represented as a direct product
\begin{equation}\label{psi}
|\Psi\rangle =\Pi_j|\psi_-\rangle_j,
\end{equation}
where $|\psi_\pm\rangle_j=(|0000\rangle_j\pm |1111\rangle_j)/\sqrt{2}$ are the two low lying states of the 8-site cluster $j$. The low-lying excited states are obtained by replacing some $|\psi_-\rangle_j$ in Eq.\ (\ref{psi}) by  $|\psi_+\rangle_j$ and have energy $4g'$ for each excited cluster.

Because the two basis states $|0000\rangle$ and  $|1111\rangle$ differ in occupancy by 4 fermions it is clear that  $\Delta_{ll'}^{(2)}=0$ for all $l,l'$ when evaluated in the ground state $|\Psi\rangle$. It is also clear that a 4-fermion expectation value $\Delta^{(4)}=\langle d^\dagger_1 d^\dagger_2d^\dagger_3 d^\dagger_4\rangle$ will be non-zero (of order one) in this ground state. Away from the $t=0$ limit the ground state does not have a simple form but, at least for small $t$, can be written approximately as a linear combination of product states constructed from $|\psi_\pm\rangle_j$. As long as this is the case the property   $\Delta_{ll'}^{(2)}=0$ and  $\Delta^{(4)}={\cal O}(1)$ continues to hold and the system must be regarded as a $4e$ superconductor. Only once $t$ grows comparable to $g'$ the ground state begins to contain a small admixture of higher energy cluster states, such as  $|0011\rangle$, possibly causing small non-zero values of $\Delta_{ll'}^{(2)}$. Nevertheless, the essential character of the topological phase remains encoded in $\Delta^{(4)}$ all the way to the phase boundary with the trivial phase. 

\vfill
\eject

\end{document}